\documentclass[a4paper,12pt]{article}
\usepackage{epsfig}
\usepackage{amssymb}
\begin{document}
\thispagestyle{empty}

\newcommand{\etal}  {{\it{et al.}}}  
\def\Journal#1#2#3#4{{#1} {\bf #2}, #3 (#4)}
\def\PRD{Phys.\ Rev.\ D}
\def\NIMA{Nucl.\ Instrum.\ Methods A}
\def\PRL{Phys.\ Rev.\ Lett.\ }
\def\PLB{Phys.\ Lett.\ B}
\def\EPJ{Eur.\ Phys.\ J}
\def\IEEETNS{IEEE Trans.\ Nucl.\ Sci.\ }
\def\CPCD{Comput.\ Phys.\ Commun.\ }

\smallskip

\bigskip
\bigskip

{\Large\bf
\begin{center}
LHC signatures of unparticles in decays of $Z^{\prime}$
\end{center}
}
\vspace{1 cm}

\begin{center}
{ G.A. Kozlov  }
\end{center}
\begin{center}
\noindent
 { Bogolyubov Laboratory of Theoretical Physics\\
 Joint Institute for Nuclear Research,\\
 Joliot Curie st., 6, Dubna, Moscow region, 141980 Russia  }
\end{center}
\vspace{1 cm}

 \begin{abstract}
 \noindent
 {We investigate the rare radiative decays of extra gauge
neutral bosons. In particular, we established the theoretical predictions for
the production of unparticle in decays of $Z^{\prime}$ - bosons.
The implementation of our prediction in the analysis at
the LHC experiments should be straightforward and lead to more precise
determination or limits of unparticle couplings and/or $Z^{\prime}$
couplings and masses.}


\end {abstract}



\section{Introduction}

\bigskip

 There is an extensive literature (see, e.g., the incomplete set of papers [1] and
 the references therein) concerning the phenomenology of
 scale-invariant hidden sector which is realized through the so-called unparticles
 in the particle physics. The unparticles (or the scale-invariant stuff) introduced
by H. Georgi in 2007 [2] with continuous mass distribution, obey the conformal (or scale) invariance.

In 1982, Banks and Zaks [3]  investigated the gauge theories containing the non-integer
numbers of Dirac fermions where  the two-loop  $\beta$-function disappeared.
In this case, one says about the non-trivial infra-red (IR) fixed point at which the theory possesses
the scale-invariant nature, and there is no chance to interpret it in terms of particles
with definite masses. The main idea is based on the following statement: at high enough values of
energy the theory contains both the fields of the Standard Model (SM) and the fields yielding
the sector with the IR point. Both of these sectors interact with each other by means
of exchange with the particles (fields) having a large mass scale $M$. At the energies below
$M$, the interaction between the fields has the form $O_{SM}\,O_{BZ}\, M^{-a}\, (a>0)$, where
$O_{SM}$ and $O_{BZ}$ are the operators of the SM fields and the Banks-Zaks (BZ) sector, respectively.
The hidden conformal sector may flow to IR fixed point scale at some scale $\Lambda < M$. In the effective theory with
energies  $\leq \Lambda$, the interaction above mentioned has the same form but composed by new set of operators
$\sim \Lambda ^{d_{BZ} - d}\,O_{SM}\,O_{U}\, M^{-a}$, where $d_{BZ}$ and $d$ mean the scale dimensions
of the operator $O_{BZ}$ and the new operator $O_{U}$ of the $U$ - unparticle, respectively.


The most interesting scale is  $\Lambda\sim$ O(TeV), at which the
dynamics of unparticles could be seen at CERN Large Hadron Collider (LHC) through the different
processes including the decays with the production of $U$- unparticles.

In the papers [2], it has been emphasized that the renormalizable interactions between the SM fields
and the fields of yet hidden conformal sector could be realized by means of explore the hidden energy
at high energy collisions and/or associated with the registration of non-integer number of
invisible particles. In this case the conformal sector described in terms of  "unparticles" does not
possess those quantum numbers which are known in the SM. For illustration purposes, we quote here the
well-known formula for the propagator $\Delta (p,m;d)$ of the scalar unparticle effectively having
the non-existing mass $m$ and the momentum  $p$ with scale dimension $1\leq d< 2$:
$$\Delta (p,m;d) = \int d^{4} x\,e^{ipx}\langle 0\vert T O_{U}(x)\,O_{U}^{+}(0)\vert 0\rangle $$
 $$ = \frac{A_{d}}{2\pi}\int^{\infty}_{m^{2}} (M^{2}-m^{2})^{d-2}\frac{i\,d M^{2}}
{p^{2}-M^{2}+i\epsilon}
 = \frac{A_{d}}{2\,\sin (d\pi)}\frac{i}{(m^{2}-p^{2} -i\epsilon)^{2-d}}.$$
This formula with the well-defined normalization constant $A_{d}$ [2] and providing the correct behavior
of the unparticle propagation in the limit  $m\rightarrow 0$, is nothing other but the propagator
of the scalar particle if $d$ assumes the canonical meaning $d=1$.

The investigation of unparticles has two directions of their study: the first one is related to
issues which carry the theoretical aspects where the main objects are associated with the interactions
of unparticles with the matter fields. One of the most important manifestations of this branch is the
coupling between the electroweak symmetry breaking through the conformal symmetry breaking
by means of input in the theory the superrenormalizable operator $\Lambda^{2-d}\,O_{U}\,H^{2}$ which
ensures the conformal symmetry breaking on the scale $(\Lambda^{2-d}\,v^{2})^{1/(4-d)}$ for
the Higgs field $H$ with its vacuum expectation value (v.e.v.) $v\simeq$ 246 GeV (see, e.g., [4,5]).
The second direction is related to the phenomenology where the main efforts are concentrated on
hunting of the unparticles in different physical processes, in particular, in decays of the
particles (scalars, pseudoscalars, gauge bosons, etc.) that might be most actual for the modern high
energy physics at the machines in operation (Tevatron, LHC) and the future accelerators (e.g., ILC).

It is a common belief the SM is extended at energies well above the weak scale.
Depending of the rank of new extended gauge group, e.g., $SO(10)$ or $E_{6}$ with the symmetry breaking
of the latter
ones to the symmetry of the SM at the certain scale, the additional neutral gauge bosons  $Z^{\prime}$
can appear, and they are associated with the additional $U^{\prime}(1)$ - symmetry.
These $Z^{\prime}$'s couple primarily to the third generation quarks.

The search for $Z^{\prime}$ at the Tevatron and at the LHC have a significant contribution in the
investigation of new physics.
The LHC will observe $Z^{\prime}$ - boson in mass region up to a few TeV.
Using the mass from the
LHC, we can test the cross-section to obtain all of the $Z^{\prime}$ -couplings.
The probability of appearance of $Z^{\prime}$ - like
objects, in particular, in the Drell-Yan channel, is discussed extensively in the literature
(see, e.g., [6] and the references therein). If such the objects are light enough this would serve as the direct message for
occurrence of new physics with a certain confirmation at the LHC even at relatively low integral
luminosity  [7,8].

Once such the particles ($Z^{\prime}$, ...) are discovered and their masses, widths, spin are defined,
it is needed to study how these new sector bosons are coupled to different SM fields, that will allow
to check the  theoretical models dealing with new physics effects. There is the hope
the LHC experiments allow to clarify the nature of a definite combination set of coupling constants of
new particles with the SM fermions by means of model-independent way. These combinations can be measured
with high precision in CERN's CMS [7] and ATLAS [8] experiments at high integral luminosity
$ \geq 100~fb^{-1}$.
The modern bounds on the
$Z^{\prime}$ mass, $m_{Z^{\prime}} > 650-900$ GeV depending on the model scheme, are already known from
experiments CDF [9] and D0 [10] at the Tevatron  (two-lepton channel). If  $Z^{\prime}$'s are coupled to
leptons, they can provide clean and spectacular signals at colliders.



The experimental channels of multi-gauge boson  production ensure the unique possibility to investigate
the anomalous triple effects of interaction between the bosons. We point out the study of
non-abelian gauge structure of the SM, and, in addition, the search for new types of interaction which,
as expected, can be evident at the energies above the electroweak scale. The triple couplings of neutral
gauge bosons, e.g., $ZZ^{\prime}\gamma$, $ZZZ^{\prime}$ etc. can be studied in pair
production at the hadron (lepton) colliders: $pp, p\bar{p} (e^{+}e^{-})\rightarrow
Z^{\prime}\rightarrow Z\gamma, ZZ,...$

Unparticle production at the LHC will be a signal that the scale where conformal invariance becomes important
to particle physics is as low as a few TeV. At this scale, the unparticle stuff sector is strongly coupled.
This requires that, somehow, a series of new reactions that involve unparticle stuff in an essential way
turn on between Tevatron and TeV energies. It will be important to understand this transition as precisely
as possible. This can be done through the LHC study of $pp \rightarrow \gamma + U$ and the identification
of the effects from $Z^{\prime}$ resonances in $pp \rightarrow fermion + antifermion$.

In this paper, we study the production of unparticle $U$ in the decays of $Z^{\prime}$ with emission
of a monophoton. In this circumstance, we have a hidden sector whose main couplings to matter fields are
through the gauge fields. Such a hidden sector can have distinct signatures at colliders.
Before to going to the concrete model, we have to make the following retreat.
First of all, we go to the extension of the Landau - Yang theorem [11,12] for the decay of a vector
particle into two vector states. Within this theorem, the decay of particle with spin-1 into two photons is
forbidden (because both outgoing particles are massless). The direct interaction between  $Z^{\prime}$ - boson
and a vector massive particle, e.g., $Z$ or $U$ - vector unparticle, accompanied by a photon, does not exist.
To the lowest order of the coupling constant $g$,  the contribution given by  $g^3$ in the decay
$Z^{\prime}\rightarrow \gamma U$ is provided mainly
by heavy quarks in the loop. What is the origin of this claim? First, it is worth to remember the
known calculation of the anomaly triangle diagram $ZZ\gamma$ [13], where the anomaly contribution result contains
two parts, one of which has no the dependence of the mass $m_{f}$ of (intermediate) charged fermions in the loop,
while the second part is proportional to  $m^{2}_{f}$. An anomaly term disappears in the case if all the fermions
from the same generation are taking into account or the masses of the fermions of each of generation
are equal to each other. The reason explaining the above mentioned note is the equality to zero of the sum
$\Sigma_{f} N^{f}_{c}\,g^{f}_{V}\,g^{f}_{A}\,Q_{f}$, where $g^{f}_{V} (g^{f}_{A})$ is the vector (axial-vector)
coupling constant of massive gauge bosons to fermions, $Q_{f}$ is the fermion charge,  $N^{f}_{c} = 3 (1)$
for quarks (leptons). The anomaly contribution for the decay $Z^{\prime}\rightarrow \gamma U$ does not
disappear due to  heavy quarks, and the amplitude of this decay is induced by the anomaly effect.
The contribution from light quarks with the mass  $m_{q}$ is suppressed as
 $m^{2}_{q}/m^{2}_{Z^{\prime}}\sim 10^{-8} - 10^{-6}$, where $m_{Z^{\prime}}$ is the mass of $Z^{\prime}$ - boson.
 Despite the decay  $Z^{\prime}\rightarrow \gamma U$ is the rare, there is a special attention to the sensitivity
 of the latter to both the top-quark and the quarks of fourth generation.

Since the photon has the only vector nature of interaction with the SM fields, the possible types
of interaction $Z^{\prime} - U - \gamma$ would be either $ V - A - V$ or  $ A - V - V$, where $V (A)$ means
the vector (axial-vector) interaction.

The anomaly triple gauge bosons contributions play an essential role in estimation of the
decay $Z^{\prime}\rightarrow Z\gamma$ normalized to
two-lepton decay or the decays with the production of quark-antiquark pairs (e.g., the ratio
$\Gamma (Z^{\prime}\rightarrow Z\gamma)/\Gamma (Z^{\prime}\rightarrow e^{+}e^{-})\sim 10^{-6}$ has been obtained
in $E_{6}$ - model [14]). Based on the extended $SU(2)_{h}\times SU(2)_{l}\times U(1)$ model in [15], the
following estimations have been done:
$\Gamma (Z^{\prime}\rightarrow Z\gamma)/\Gamma (Z^{\prime}\rightarrow l\bar l)\sim 10^{-5} (\mu^{+}\mu^{-}),
\sim 10^{-4}- 10^{-3} (\tau^{+}\tau^{-})$, $\Gamma (Z^{\prime}\rightarrow Z\gamma)/\Gamma (Z^{\prime}\rightarrow q\bar q)
\sim 10^{-6}~ (q: b,top)$  with the account of the fourth generation quarks.
Note, that the papers on the $U$ - unparticle production in radiative decays  $H \rightarrow \gamma U$ and
$Z \rightarrow \gamma U$  are published in [16] and [17], respectively.

The main reason of the study in this work is related to the fact that
the $Z^{\prime}$ couples to quarks both with a vector and axial-vector coupling, and it follows that
in radiative decays one can reach all spin-parity final states. This suggest, then, that it might be
possible to use these mean to reach the unparticles, in particular, to estimate
the contribution from the $Z^{\prime}$ in its rare radiative decays to $U$ - unparticle and
to take into account the possible contributions from
 up - and down - quarks of the fourth generation with the masses  $> 258-268$ GeV (see, e.g., the data by CDF [18]
 and the references in [19]). Note, the fourth generation quarks lead to new dimensionless effective
constants $\lambda_{1},\, \bar\lambda_{1},\,c_{v},\, a_{v}$ in the interactions of the form
\begin{equation}
\label{eq1}
 \lambda_{1}\frac{c_{v}}{\Lambda^{d-1}}\,\bar q\,\gamma_{\mu}\,q\,O_{U}^{\mu},\,\,
\bar\lambda_{1}\frac{a_{v}}{\Lambda^{d-1}}\,\bar q\,\gamma_{\mu}\,\gamma_{5}\,q\,O_{U}^{\mu},
\end{equation}
where $O_{U}^{\mu}$ being the operator of the unparticle (spin-1) on the scale $\Lambda$ with dimension $d$.

In Sec. 2, we will present the model. Section 3 focuses on the calculations of the photon energy spectrum,
the branching ratio of the decay $Z^{\prime}\rightarrow\gamma U$ and the cross section of the process
$pp\rightarrow Z^{\prime}\rightarrow\gamma U$. In Sec. 4, we give the experimental constraints for
$U$-unparticle observable. Our conclusions are presented in Sec. 5.

\section{Model}
Let us consider the following interaction Lagrangian density

\begin{equation}
\label{eq2}
-L = g_{Z^{\prime}}\sum_{q} \bar q (v^{\prime}_{q}\gamma^{\mu} -
a^{\prime}_{q}\gamma^{\mu}\gamma_{5}) q Z^{\prime}_{\mu} +
\frac{1}{\Lambda^{d-1}}\sum_{q} \bar q (\lambda_{1} c_{v}\gamma^{\mu} - \bar\lambda_{1} a_{v}\gamma^{\mu}\gamma_{5})
q O_{U}^{\mu},
\end{equation}
where $g_{Z^{\prime}} = (\sqrt {5b/3}\,s_{W}\,g_{Z})$ is the gauge constant of $U^{\prime}(1)$ group
(the coupling constant of $Z^{\prime}$ with a quark $q$) with the group factor $\sqrt {5/3}$,
$b\sim O(1)$, $g_{Z}=g/c_{W}$; $s_{W}(c_{W})= \sin\theta_{W} (\cos\theta_{W})$, $\theta_{W}$ is the angle of weak
interactions (often called as Weinberg angle); $v^{\prime}_{q}$ and $a^{\prime}_{q}$ are generalized vector and
the axial-vector  $U^{\prime}(1)$ -charges, respectively, which are $Z^{\prime}$-pattern model-dependent, а
for example, in the frame of $E_{8}\times E_{8},\,E_{6}, ...$ groups. These latter charges are dependent on both (joint)
gauge group  and the Higgs representation which is responsible for the breaking of initial gauge group to
the SM one; $\lambda_{1},\, \bar\lambda_{1}$ are defined in (\ref{eq1}).
Actually, the second term in (\ref{eq2}) is identical to the first one up to the $d$-dimensional factor
$\Lambda^{1-d}$.
In the model, we assume
 $O{_{\mu}}_{U}$ is a non-primary operator derived by $O{_{\mu}}_{U}(x) \sim \partial_{\mu} S(x)$ through the
 pseudo-Goldstone field of conformal symmetry $S(x)$ - pseudo-dilaton mode. The  scalar field $S(x)$ serves
 as a conformal compensator with continuous mass. Because the conformal sector is strongly coupled, the
 mode $S(x)$ may be one of new states accessible at high energies.
 $O{_{\mu}}_{U}$ has both the vector and the axial-vector couplings to quarks in the loop.

It is known that both $Z$ and $Z^{\prime}$ are not the mass eigenstates. There is the $Z$-$Z^{\prime}$ mixing
(with mixing angle $\varphi$)
which leads to rotation of the neutral sector to the physical states $Z_{1}$ and $Z_{2}$ with the masses $m_{1}$
and $m_{2}$, respectively:
$$\tan^{2}\varphi = \frac{m^{2}_{Z}-m^{2}_{1}}{m^{2}_{2}-m^{2}_{Z}}.$$
Within its small absolute value [15], the angle $\varphi$ does not play an essential role in the calculations.
It turns out, one
can identify  $Z$ and $Z^{\prime}$ with the physical neutral states. Actually, $m_{1}\simeq m_{Z} << m_{2}\simeq
m_{Z^{\prime}}$.

We consider the model [20,21] containing the $Z_{\chi}$- boson on the scale
$O(1~TeV)$ in the frame of the symmetry based on the $E_{6}$ effective gauge group. The group $U(1)_{\chi}$
arises in the following consequence of the symmetry breaking of the parent group $E_{6}$:
$E_{6}\rightarrow SO(10)\times U(1)_{\psi}\rightarrow SU(5)\times U(1)_{\chi} \times U(1)_{\psi}$.
We suppose that $SO(10)$ breaks to  $SU(5)\times U(1)_{\chi}$ at the same scale where $SU(5)$ breaks to
the SM group $SU(3)_{c}\times SU(2)_{L}\times U(1)_{Y}$ with respective coupling constants $g_{s}$, $g$, $g^{\prime}$.
The coupling constant of $U(1)_{\chi}$ has the form $g_{\chi} = \sqrt{5/3}\,e/c_{W}$.

The amplitude of the decay$Z^{\prime}\rightarrow U\gamma$, where the coupling $Z^{\prime}\, U\,\gamma$ is supposed to
be extended by the intermediate loop containing the quarks $q$, has the form:
\begin{equation}
\label{eq4}
Am(z_{u},z_{q}) = \frac{e^{2}}{c_{W}}\sqrt{\frac{5}{3}}\frac{3}{\Lambda^{d-1}}
\sum_{q} e_{q}  \left (\lambda_{1} c_{v} a^{\prime}_{q} +
\bar\lambda_{1}\, a_{v}\,v^{\prime}_{q}\right ) I(z_{u},z_{q})
\end{equation}
with $z_{u}=P^{2}_{U}/m^{2}_{Z^{\prime}}$,
$z_{q}=m^{2}_{q}/m^{2}_{Z^{\prime}}$ for the momentum $P_{U}$ of $U$ - unparticle
and the quarks  $q$ (in the loop) with the mass $m_{q}$.
We deal with the following expression for $I(z_{u},z_{q})$:
\begin{equation}
\label{eq44}
I=\frac{1}{1-z_{u}}\left \{\frac{1}{2}+\frac{z_{q}}{1-z_{u}}\left [F(z_{q}) -
F\left (\frac{z_{q}}{z_{u}}\right)\right]-\frac{1}{2(1-z_{u})}\left [G(z_{q}) -
G\left (\frac{z_{q}}{z_{u}}\right)\right]\right\}
\end{equation}
adopted for the decay  $Z^{\prime}\rightarrow U\gamma$ taking into account the results obtained in
[14] and [17].
For heavy quarks, $m_{q} > 0.5 \,m_{Z^{\prime}}$, the functions $F(x)$ and $G(x)$  in (\ref{eq44})
have the following forms [14]:
\begin{equation}
\label{eq45} F(x)= - 2 {\left (\sin^{-1}\sqrt {\frac{1}{4x}}\right
)}^{2},\,\, G(x) = 2\sqrt {4x-1}\,\sin^{-1} \left (\sqrt
{\frac{1}{4x}}\right ),
\end{equation}
while for light quarks ($m_{q} < 0.5 \,m_{Z^{\prime}}$), one has to use the formulas:
\begin{equation}
\label{eq46}
F(x)= \frac{1}{2}{\left (\ln\frac{y^{+}}{y^{-}}\right)}^{2} + i\pi\ln\frac{y^{+}}{y^{-}}
-\frac{\pi^{2}}{2},\,\,\,\,
G(x) = \sqrt{1-4x} \left (\ln{\frac{y^{+}}{y^{-}}} + i\pi\right ),
\end{equation}
where $y^{\pm} = 1 \pm\sqrt {1 -4\,x}$.
To get (\ref{eq46}), we have used the complex continuation applied to (\ref{eq45}): $\sin^{-1}\sqrt {1/4x}\rightarrow
i(arccosh \sqrt {1/4x} + i\,\pi/2) = i(\ln y^{+}/y^{-} + i\,\pi)/2$ and $\sqrt {4\,x -1} \rightarrow - i\,\sqrt {1-4\,x}$.
The variable $z_{u}$ is related to the energy $E_{\gamma}$ of the photon as $z_{u} = 1 - 2\,E_{\gamma}/m_{Z^{\prime}}$.
In the frame of the  $Z_{\chi}$ - model, we choose $v^{\prime}_{up} = 0,\, a^{\prime}_{up} = \sqrt{6}\,s_{W}/3,\, v^{\prime}_{down} =
2\,\sqrt{6}\,s_{W}/3,\, a^{\prime}_{down} = -\sqrt{6}\,s_{W}/3$ for $up$ - and $down$ - quarks. Actually,
 the account of the only light quarks leads to the zeroth result for the amplitude (\ref{eq4}).
The contribution $\sim \bar\lambda_{1}\,a_{v}\,v^{\prime}_{q}$ is nonzero for the only $b$ - quarks and down - quarks
of fourth generation. We emphasize that the nonvanishing result for the amplitude $Z^{\prime}\rightarrow U\gamma$ is
the reflection of the anomaly contribution due to the presence of heavy quarks.


\section{Photon energy spectrum, branching ratio and cross section}

In the decay $Z^{\prime}\rightarrow U\gamma$, the unparticle can not be identified with the definite invariant mass.
$U$- unparticles stuff possesses by continuous mass spectrum, and can not be in the rest frame (there is the similarity
to the massless particles). Since the unparticles are stable (and do not decay), the experimental signal of their
identification could be looking via the hidden (missing) energy and/or the measurement of the momentum distributions in the case
when the $U$- unparticle is produced in $Z^{\prime}\rightarrow \gamma\,U$ or $Z^{\prime}\rightarrow \bar f f\,U$.

The differential distribution of the decay width а $\Gamma(Z^{\prime}\rightarrow U\gamma)$ over the variable
$z_{u}$ looks like (see also [17]):
\begin{equation}
\label{eq5}
\frac{d\Gamma}{d z_{u}} = \frac{1}{2m_{Z^{\prime}}}\, \sum {\vert M\vert}^{2}\frac{A_{d}}
{16\,\pi^{2}}\left (m^{2}_{Z^{\prime}}\right )^{d-1}\, z^{d-2}_{u} (1-z_{u}),
\end{equation}
where
\begin{equation}
\label{eq6}
 \sum {\vert M\vert}^{2} = \frac{1}{6\,\pi^{4}}\, z_{u}\,(1-z_{u})^{2}\,(1+z_{u})\,
 {\vert Am(z_{u},z_{q})\vert}^{2}\, m^{2}_{Z^{\prime}},
\end{equation}
and
$$A_{d} = \frac{16\,\pi^{5/2}}{(2\,\pi)^{2d}}\, \frac{\Gamma(d+1/2)}{\Gamma(d-1)\,\Gamma(2d)}.$$

One of the requirements applied to the amplitude in (\ref{eq6}) is that it
disappears in case of "massless" unparticle (Landau-Yang theorem), and when $z_{u} =1$.

Some bound regimes in (\ref{eq5}) may be both useful and instructive for further investigation.
For this, we consider the quark-loop couplings in the amplitude (\ref{eq4}) as the sum
of the contributions given by light quarks $q$ and heavy ones $Q$
(the contribution from the quarks of 4-th generation is also possible):
$$\sum_{q} e_{q}  \left (\lambda_{1} c_{v} a^{\prime}_{q} +
\bar\lambda_{1}\, a_{v}\,v^{\prime}_{q}\right ) I(z_{u},z_{q}) +
\sum_{Q} e_{Q}  \left (\lambda_{1} c_{v} a^{\prime}_{Q} +
\bar\lambda_{1}\, a_{v}\,v^{\prime}_{Q}\right ) I(z_{u},z_{Q}).$$
For light quarks, $z_{q} << 1$, there are no $q$ - mass dependence in the one-loop function
$I(z_{u},z_{q})$ in the limit $m_{q}\rightarrow 0$
$$I(z_{u},z_{q}) \simeq \frac{1}{2(1- z_{u})}\,\left [1 - 4\,\frac{z_{q}}{z_{u}}\left (
\frac{10}{3} + i\,\pi\right )\right ]. $$
On the other hand, if the quarks inside the loop become heavy enough, $z_{Q} > 1$, we estimate the
following  function $I(z_{u},z_{Q})$
$$I(z_{u},z_{Q}) \simeq \frac{1}{24\,z_{Q}}\left [1+\left (\frac{2}{7}\right )^{2}\frac {
1+ z_{u}}{1 - z_{u}}\right ] -\frac{1}{576 z^{2}_{Q}}\frac{1 + z_{u} + z^{2}_{u}}{1-z_{u}}, $$
which is very small in the limit $z_{Q} >> 1$.

For the unparticle spectrum
at low rates of $z_{u}$, the only heavy quarks $Q$ with the masses $m_{Q} > 0.5 m_{Z^{\prime}}$
are responsible for the production of $U$ -unparticles and the distribution ${d\Gamma}/{d z_{u}}$
has the form
\begin{equation}
\label{eq8}
  \lim_{z_{u}\rightarrow 0} \frac{d\Gamma}{d z_{u}} \simeq
  \frac{5A_{d}}{(2\pi)^{6}}\left (\frac{e^{2}}{c_{W}}\right )^{2}
  \left (\frac{z_{u}}{\Lambda ^{2}}\right )^{d-1}
 \left (m^{2}_{Z^{\prime}}\right )^{d-1/2}{\left\vert \sum_{Q} e_{Q} (\lambda_{1}c_{v}a^{\prime}_{Q} +
  \bar\lambda_{1}a_{v}v^{\prime}_{Q}) I(z_{Q})\right\vert}^{2},
\end{equation}
where $I(z_{Q})=0.5 [1-G(z_{Q})] + z_{Q}\cdot F(z_{Q})$ for $z_{Q} > 1/4$ and $d\neq 1$ .
In this case, the photon energyа
$E_{\gamma} = m_{Z^{\prime}} (1- m_{z_{u}})/2$ in the limit $z_{u}\rightarrow 0$ gets its finite value.
The limit ${z_{u}\rightarrow 1}$ is trivial and we do not consider this.


Within the fact of the combination $\lambda_{1} c_{v} a^{\prime}_{q} + \bar\lambda_{1}\, a_{v}\,v^{\prime}_{q}$
in (\ref{eq4}) with the effective constants of unparticles   $\lambda_{1} c_{v}$ and
$\bar\lambda_{1}\, a_{v}$, the decay amplitude of $Z^{\prime}\rightarrow \gamma U$ does not disappear when the
summation on all the quarks degree of freedom is performed.

In Table 1, the monophoton energy distribution $E_{\gamma}^{-1}\,d\Gamma/dz_{u}$ is tabulated for a
range of photon energy  $E_{\gamma}$ = 0 - 450 GeV and various choices of $d$. For simplicity, we use the
$E_{\chi}$-model assuming the
flavor blind universality $\lambda_{1}\,c_{v} = \bar\lambda_{1}\,a_{v} =1$ for all three generation quarks,
$\Lambda$ and $m_{Z^\prime}$ are set to be 1 TeV each.

{ \it {\bf Table 1.}
Energy distribution $E_{\gamma}^{-1}\,d\Gamma/dz_{u}\times 10^{8}$ for  $E_{\gamma}$ = 0 - 450 GeV,
 depending on
$d= 1.1 - 1.9$; $\lambda_{1}\,c_{v} = \bar\lambda_{1}\,a_{v} =1$, $\Lambda$ = 1 TeV, $m_{Z^{\prime}} =$ 1 TeV.}

\begin{center}
\begin{tabular}{ c c c c c c c c} \hline
\hline
$E_{\gamma}, GeV $ &$ 0$ &$ 50$ & $ 100$ & $200$ & $300$ & $400$ & $ 450$ \\[4 mm]
\hline
$d = 1.1 $ & 5.70 & 5.41 & 5.10 & 4.40 & 3.74  & 2.90 & 2.53\\[0.9 mm]
$d = 1.3$ & 7.10 & 6.64 & 6.03 & 4.92 & 3.83 &  2.76 &  1.92 \\[0.9 mm]
$d = 1.5 $ & 4.42 & 4.05 & 3.50 & 2.70 & 1.92 &  1.25 & 0.74  \\[0.9 mm]
$d = 1.7$ & 2.24 & 1.90 & 1.73 & 1.23 &  0.80 &  0.40 & 0.24  \\[0.9 mm]
$d = 1.9$ & 0.80 & 0.70 & 0.64 & 0.40 & 0.30 & 0.11 &  0.06  \\[0.9 mm]
\hline
\end{tabular}
\end{center}
The sensitivity of the scale dimension $d$ to the energy distribution is evident, and
$ E_{\gamma}^{-1}\,d\Gamma/dz_{u}$ decreases with the photon energy $E_{\gamma}$.

The $U$-unparticle could behave as a very broad vector boson since its mass could be distributed
over a large energy spectrum. The production cross-section into each energy bin could be much smaller
than in the case where a SM vector boson has that particular mass. This may be the reason why we have not yet
seen the  $U$-unparticle trace in the experiment.

In the appropriate approximation when the relation between the total decay width $\Gamma_{Z^{\prime}}$ and
$ m_{Z^{\prime}}$ is small, the contribution to the cross section of the process
$pp\rightarrow Z^{\prime}\rightarrow \gamma\, U$ can be separated into $Z^{\prime}$ production cross section
$\sigma (pp\rightarrow Z^{\prime})$ and the relevant branching fraction of the $Z^{\prime}$ - boson:
$\sigma (pp\rightarrow Z^{\prime}\rightarrow \gamma\, U) =  \sigma (pp\rightarrow Z^{\prime})\cdot
B(Z^{\prime}\rightarrow \gamma\, U)$.
The $Z^{\prime}$ state can be directly produced at a hadron collider via the $\bar q q\rightarrow Z^{\prime}$
subprocesses, for which the cross section in the case of infinitely narrow $Z^{\prime}$ is given by
\begin{equation}
\label{eq661}
\sigma (\bar {q}q\rightarrow Z^{\prime}) = k_{QCD}\,\frac{4\,\pi^{2}}{3}\,\frac{\Gamma (Z^{\prime}\rightarrow \bar {q}q)}
{m_{Z^{\prime}}}\,\delta (\hat s - m_{Z^{\prime}}^{2}),
\end{equation}
where $k_{QCD}\simeq$ 1.3 represents the enhancement from higher order QCD processes.
Conservation of the energy-momentum implies that the invariant mass of $Z^{\prime}$ is equal to the
parton center-of-mass energy $\sqrt{\hat {s}}$, where $\hat {s} = x_{1}x_{2}s$ depends of the
fractions, $x_{1}$ and $x_{2}$, of those momenta carried by partons that initiate the process
$\bar {q} q \rightarrow Z^{\prime}$.
The decay width $\Gamma (Z^{\prime}\rightarrow \bar {q}q)$ is
$$\Gamma (Z^{\prime}\rightarrow \bar {q}q) =  \frac{G_{F}\,m_{Z}^{2}}{6\pi\,\sqrt{2}}\,N_{c}\,m_{Z^{\prime}}\,
\sqrt {1- 4\,z_{q}}\,\left [\left (v^{\prime}_{q}\right )^{2} (1+ 2\,z_{q}) +  \left (a^{\prime}_{q}\right )^{2}(1- 4\,z_{q})\right ], $$
where $G_{F}$ is the Fermi coupling constant. In the narrow width approximation, the cross section (\ref{eq661}) reduces to
 ($z_{q} << 1$)
\begin{equation}
\label{eq666}
\sigma (\bar {q}q\rightarrow Z^{\prime}) \simeq k_{QCD}\,\frac{2\,a}{3}\,\frac{G_{F}}{\sqrt{2}}
\left (\frac{m_{Z}}{m_{Z^{\prime}}}\right )^{2}\frac{\left [\left (v^{\prime}_{q}\right )^{2} +
\left (a^{\prime}_{q}\right )^{2}\right ]}{(\hat {s}/m_{Z^{\prime}}^{2} -1 )^{2} + a^{2}},
\end{equation}
where $a = \Gamma_ {Z^{\prime}}/m_{Z^{\prime}}$.

For quarks obeying the condition $0 \leq  z_{q} \leq 1/4$ the branching ratio
for $Z^{\prime}\rightarrow\gamma\,U $ decay,
$B (Z^{\prime}\rightarrow\gamma\,U) = \Gamma (Z^{\prime}\rightarrow\gamma\,U)/
\Gamma_{Z^{\prime}}$, does not disappear even at $z_{q} = 0$
\begin{eqnarray}
\label{eq91}
 B (Z^{\prime}\rightarrow\gamma\,U) = \frac{5\,A_{d}}{a\, c_{W}^{2}}\left (\frac{\alpha}{2\,\pi^{2}}\right )^{2}
  \left (\frac{m_{Z^{\prime}}^{2}}{\Lambda^{2}}\right )^{d-1} \cr
  {\left\vert \sum_{q} e_{q} \,(\lambda_{1}\,c_{v}a^{\prime}_{q} +
  \bar\lambda_{1}\,a_{v}v^{\prime}_{q})\right\vert}^{2}\left [\frac{1}{2\,d\,(d+2)} -\frac{20}{3}\frac{\Gamma
  \left (\frac{d-1}{2}\right )}{(d+1)\,\Gamma \left (\frac{d+1}{2}\right )}\,z_{q}\right ].
\end{eqnarray}
In Table 2, $B (Z^{\prime}\rightarrow\gamma\,U)$  is tabulated with the assumption of
$\lambda_{1}\,c_{v} = \bar\lambda_{1}\,a_{v} =1$ for all three generation quarks,
$\Lambda$ is set to be 1 TeV, the range of  $d$ is chosen as = 1.1 - 1.9 for the definite spectrum of
$m_{Z^{\prime}} =$ 0.5; 0.7; 0.9; 1.1; 2.0; 3.0 TeV. The total decay
width $\Gamma_{Z^{\prime}}$ of $Z^{\prime}$-boson is chosen in the framework of the
Sequential SM ($Z^{\prime}_{SSM}$) , where
the ratio $\Gamma_{Z^{\prime}}/m_{Z^{\prime}}$
has the maximal value $ a= 0.03$ among the Grand Unification Theories (GUT) inspired $Z^{\prime}$
models (as a review, see, for example, the paper by P. Langacker in [6]) .\\

{\it {\bf Table 2.} Branching ratio $B (Z^{\prime}\rightarrow\gamma\,U)\times 10^{7}$ (\ref{eq91}) for
$\lambda_{1}\,c_{v} = \bar\lambda_{1}\,a_{v} =1$, $\Lambda =1$ TeV depending on
$d= 1.1 - 1.9$ and $m_{Z^{\prime}} =$ 0.5; 0.7; 0.9; 1.1; 2.0; 3.0 TeV with
$\Gamma_{Z^{\prime}}$ given by the $Z^{\prime}_{SSM}$-model ($a = 0.03$)}

\begin{center}
\begin{tabular}{ c c c c c c c } \hline
\hline
$m_{Z^{\prime}}, TeV $ &$ 0.5$ &$ 0.7$ & $ 0.9$ & $1.1$ & $2.0$ & $3.0$ \\[4 mm]
\hline
$d = 1.1 $ & 2.39 & 2.56 & 2.69 & 2.80 & 3.16  & 3.42\\[0.9 mm]
$d = 1.3$ & 1.78 & 2.18 & 2.54 & 2.88 & 4.11 &  5.23\\[0.9 mm]
$d = 1.5 $ & 0.68 & 0.95 & 1.22 & 1.50 & 2.72 &  4.08\\[0.9 mm]
$d = 1.7$ & 0.21 & 0.34 & 0.49 & 0.65 &  1.50 &  2.64\\[0.9 mm]
$d = 1.9$ & 0.053 & 0.093 & 0.15 & 0.21 & 0.63 & 1.31\\[0.9 mm]
\hline
\end{tabular}
\end{center}

We find the smooth increasing of $B (Z^{\prime}\rightarrow\gamma\,U)$ with
$m_{Z^{\prime}}$ and its decreasing with the dimension $d$.\\
In Table 3, the cross section $\sigma (\bar q q\rightarrow Z^{\prime}\rightarrow \gamma\, U)$
is tabulated in the case of up-quarks annihilation, where $x_{1}\sim x_{2}\sim \sqrt {x_{min}}$,
$x_{min} = m_{Z^{\prime}}^{2}/s$; the range of  $d$ is chosen as = 1.1 - 1.9 for
$m_{Z^{\prime}} =$ 0.5; 0.7; 0.9; 1.1; 2.0; 3.0 TeV;  $ a= 0.03$.

{ \it {\bf Table 3.} Cross section $\sigma (\bar q q\rightarrow Z^{\prime}\rightarrow \gamma\, U)\times 10^2, fb$
with the assumption of up-quarks annihilation, where $d$ = 1.1 - 1.9,
$m_{Z^{\prime}} = 0.5; 0.7; 0.9; 1.1; 2.0; 3.0 $TeV,  $ a= 0.03$.}

\begin{center}
\begin{tabular}{ c c c c c c c } \hline
\hline
$m_{Z^{\prime}}, TeV $ &$ 0.5$ &$ 0.7$ & $ 0.9$ & $1.1$ & $2.0$ & $3.0$ \\[4 mm]
\hline
$d = 1.1 $ & 11.20 & 5.90 & 4.00 & 2.70 & 0.92  & 0.44\\[0.9 mm]
$d = 1.3$ & 8.37 & 5.01 & 3.81 & 2.80 & 1.19 &  0.70\\[0.9 mm]
$d = 1.5 $ & 3.20 & 2.19 & 1.83 & 1.45 & 0.79 &  0.53\\[0.9 mm]
$d = 1.7$ & 0.99 & 0.78 & 0.74 & 0.63 &  0.44 &  0.34\\[0.9 mm]
$d = 1.9$ & 0.25 & 0.21 & 0.21 & 0.20 & 0.18 & 0.18\\[0.9 mm]
\hline
\end{tabular}
\end{center}
For $100~fb^{-1}$ luminosity at the LHC, we find
for a 0.9 TeV $Z^{\prime} $ and with up-type quarks annihilation, a small number of events,
corresponding to 4 signal events at $d = 1.1$, while at  $d = 1.9$ this number does not exceed
the single one. In the case of down-type quarks, the events estimation is five times more which
is more optimistically to be observed at the LHC.

It is known that for the pole mass, e.g., $m_{Z^{\prime}} = $ 1 TeV, the invariant
di-lepton mass resolutions for Drell-Yan processes $pp\rightarrow \bar l\, l$ for
both ATLAS [8] and CMS [7] detectors are about 2 $\%$ for $e^{+}e^{-}$ - and
twice more in the case of $\mu^{+}\mu^{-}$ - final pairs. It means that for GUT inspired $Z^{\prime}$,
LHC will be at the threshold or even not be able to measure the $\bar l\,l$ - pair production
via the  $Z^{\prime}$ - boson decay. On the other hand, within the aim to explore the
$U$ - unparticles, the parent particle, $Z^{\prime}$, can also be in the state containing the
continuously distributed mass. Such a  $Z^{\prime}$ - boson looks like, e.g., the $E_{6}$ $Z^{\prime}$
with some nonzero internal decay $\Gamma_{Z^{\prime}}^{int}$ width determined by the spectral
function $\rho (t)$  in the following approximate equality [22]:
$$\int_{0}^{\infty} \frac{\rho (t)\,dt}{p^{2} - t- i\,\epsilon} \simeq \frac{1}
{p^{2} - m_{Z^{\prime}}^{2} + i\,m_{Z^{\prime}}\, \Gamma_{Z^{\prime}}^{int}},$$
where
$$\rho(t) = \frac{1}{\pi}\frac{m_{Z^{\prime}}\,\Gamma_{Z^{\prime}}^{int}}{(t - m_{Z^{\prime}}^{2})^{2} +
(\Gamma_{Z^{\prime}}^{int})^{2}\,m_{Z^{\prime}}^{2}}. $$
The relatively large ratio $ \Gamma_{Z^{\prime}}^{int} / m_{Z^{\prime}}$ is bigger than the
$\bar l\,l$ invariant mass resolution, and, then, the discovery of broad heavy vector resonance
will clarify the internal structure of  $Z^{\prime}$.

\section{$U$- observable and experimental constraints}

In some sense, the unparticle sector carries the SM features in terms of operators.
It means that the hidden sector could be strongly constrained by existing experimental
data. One of the important and practical implications for unparticle
phenomenon in the framework of experimental constraints is the analysis of the operator
form (already mentioned schematically in the Introduction):
\begin{eqnarray}
\label{eq10}
 const\, \frac{\Lambda^{d_{BZ} - d}}{M^{d_{BZ} - 2}}\,\vert H\vert^{2}\,O_{U}
\end{eqnarray}
containing the Higgs field $H$ in the IR with dimension $d$ towards the breaking
of the conformal invariance. Within the Higgs v.e.v.
requirement, the theory becomes nonconformal below the scale
$$\tilde\Lambda = \left (\frac{\Lambda^{d_{BZ} - d}}{M^{d_{BZ} - 2}}\,v^{2}\right )^
{\frac{1}{4-d}} < \Lambda, $$
where $U$-unparticle sector becomes as known particle one. For practical (experimental)
consistency we require $\tilde\Lambda < \sqrt {s}$. It implies that unparticle physics phenomena
can be seen at high energy experiment with the energies
$$s > \left (\frac{\Lambda^{d_{BZ} - d}}{M^{d_{BZ} - 2}}\,v^{2}\right )^
{\frac{2}{4-d}}$$
even when $d\rightarrow d_{BZ}$. Note, that any observable involving operators
$O_{SM}$ and $O_{U}$ in (\ref{eq10}) will be given by the operator
$$\hat o = \left (\frac{\Lambda^{d_{BZ} - d}}{M^{d_{BZ} + n - 4}}\right )^
{2}\, s^{ d + n -4}, $$
where $ n$ is the dimension of the SM operator. Then, the observation of the unparticle
sector is bounded by the minimal energy
\begin{eqnarray}
\label{eq11}
s > {\hat o}^{\frac{1}{n}}\, M^{2}\, \left (\frac{v}{M}\right )^{\frac{4}{n}}.
\end{eqnarray}
No both $d$- and $d_{BZ}$ - dimensions we have in the lower bound (\ref{eq11}). The main
model parameter is the mass $M$ of heavy messenger. If the experimental deviation from
the SM is detected at the level of the order $\hat o \sim 1 \%$ at $n = 4$, the
lower bound on $\sqrt {s}$ would be from 0.9 TeV to 2.8 TeV for $M's$ running from
10 TeV to 100 TeV, respectively. Thus, both the Tevatron and the LHC are the ideal
colliders where the unparticle physics can be tested well.

\section{Conclusions}
In conclusion, we have studied the decay of extra neutral gauge boson $Z^{\prime}$
into vector $U$-unparticle and
a single photon. Both vector and axial-vector couplings to quarks (including the heavy
quarks of the 4th generation) play a significant role.
A nontrivial scale invariance sector of dimension $d$ may give rise to peculiar missing
energy distributions in $Z^{\prime}\rightarrow\gamma\,U$ that can be treated at the LHC.
The energy distribution for $pp\rightarrow Z^{\prime}\rightarrow \gamma\, U$ can discriminate $d$.
The branching ratio $B (Z^{\prime}\rightarrow\gamma\,U)$ is small and at best of the order of $10^{-7}$
for small scale dimension $d = 1.1$. For larger $d$, the branching ratio is at least smaller by one order
of the magnitude.
For $100~fb^{-1}$ luminosity at the LHC, we find the required production range for $\gamma\,U$ is around
(4 - 20) signal events at $d = 1.1$ for a 0.9 TeV $Z^{\prime}$, while for larger values of $d$ the events number decreases sharply.

Unless the LHC can collect a very large sample of $Z^{\prime}$, detection of $U$- unparticles
through $Z^{\prime}\rightarrow \gamma\,U$ would be quite challenging.
At the LHC or even at the ILC, we
expect the beam energies which can be run at the $Z^{\prime}$-pole, and with a large number of   $Z^{\prime}$'s
the branching ratio $B (Z^{\prime}\rightarrow \gamma\,U)$ as low as $10^{-8} - 10^{-7}$ can be tested.

For the case when $Z^{\prime}$ - boson has continuously distributed mass, the branching ratio has an
additional suppression factor due to nonzero $\Gamma_{Z^{\prime}}^{int}$. The experimental estimation
of $B (Z^{\prime}\rightarrow \gamma\,U)$ will provide us with the quantity of  $\Gamma_{Z^{\prime}}^{int}$.

The implementation of our prediction in the LHC analyses should be straightforward and
lead to more precise determination or limits of unparticles couplings and/or $Z^{\prime}$ couplings and masses.
We have shown numerical results for $Z^{\prime}$ -bosons associated with the $Z_{\chi}$ - model. The
calculations are easily applicable to other extended gauge models, e.g., Little Higgs scenario
models, Left-Right Symmetry Model, Sequential SM.


We have ignored issues of efficiencies and backgrounds; these must of course be included in a data
analyses.

The new phenomena like the  $Z^{\prime}$ - bosons with the mass $\sim $O(1 TeV)
could be discovered at the LHC in 2011 run, optimistically, at 100 $fb^{-1}$ and $\sqrt s \sim $ 8-10 TeV.
When physics (beyond the SM) of the $Z^{\prime}$ - bosons becomes  a reality, it will be much
clearer to the physics community what the unparticle stuff is for and why it is needed to be
incorporated in new physics.
If the model of new couplings between  $Z^{\prime}$ and $U$-unparticles that
we are discussing in this paper is correct, the first signs of that new physics will be discovered
at the LHC.




\end{document}